\documentclass[useAMS,usenatbib]{mn2e}

\usepackage{graphicx,color,afterpage}
\usepackage[total={17.8cm,24.0cm},centering]{geometry}
\usepackage{times,subfigure,rotating}
\def\hi{H{\small I}}

\def\CeightO{C$^{18}$O}

\def\arcsec{$^{\prime \prime}$}
\definecolor{Mygrey}{gray}{0.75}

\newcommand{\ltsimeq}{\raisebox{-0.6ex}{$\,\stackrel{\raisebox{-.2ex}{$\textstyle <$}}{\sim}\,$}}

\newcommand{\co}[1]{$^{#1}$CO}

\usepackage[compact]{titlesec}
\titlespacing{\section}{0pt}{*2}{*1}

\title[Systematic variation of the \co{12} / \co{13} ratio with $\Sigma_{\rm SFR}$]{Systematic variation of the \co{12} / \co{13} ratio as a function of star-formation rate surface density} 
\author[Timothy A. Davis et al.]{Timothy A. Davis$^{1,2}$\thanks{E-mail:\texttt{tdavis@eso.org}}
 \vspace{0.4cm}\\
\parbox{\textwidth}{$^{1}$European Southern Observatory, Karl-Schwarzschild-Str. 2, 85748 Garching, Germany\\
$^{2}$Centre for Astrophysics Research, University of Hertfordshire, Hatfield, Herts AL1 9AB, UK
}}

\begin{document}

\date{Accepted 2014 September 05. Received 2014 September 04; in original form 2014 June 05}

\pagerange{\pageref{firstpage}--\pageref{lastpage}} \pubyear{2012}

\maketitle

\label{firstpage}

\begin{abstract}
We show that the \co{12}/\co{13}  {intensity} ratio in nearby galaxies varies systematically as a function of the star formation rate surface density ($\Sigma_{\rm SFR}$) and gas surface density. The same effect is observed in different transitions, and in the \co{12}/\CeightO\ ratio, while the \co{13}/\CeightO\ ratio appears to remain constant as a function of $\Sigma_{\rm SFR}$. We discuss the cause of these variations, considering both changes in the physical state of the gas, and chemical changes that lead to abundance variations. 
We used the observed correlations with \CeightO\ to suggest that abundance variations are unlikely to be causing the systematic trend observed with $\Sigma_{\rm SFR}$, and thus that the mean gas temperature and/or velocity dispersion are systematically higher in higher star-formation rate surface density regions. We present the best fitting relations between $\Sigma_{\rm SFR}$ and the \co{12}/\co{13} and \co{12}/\CeightO\ ratios, and discuss how this effect can help us predict CO isotope emission from galaxies across the known universe.

\end{abstract}


\section{Introduction}
\label{intro}
Improving our knowledge of the build-up of stellar mass in the universe is crucially linked to better understanding the behaviour of baryons in the gas phase, especially in the cold atomic and molecular phases that are tightly linked to star-formation. The most abundant hydrogen component of atomic gas is directly observable in the local universe, thanks to the 21cm transition of \hi. Tracing molecular H$_2$, however, requires the use of proxy molecules (such as carbon monoxide) which have suitable transitions in the millimetre regime, and which are excited even in the cold conditions found in typical molecular clouds. 

In the ALMA (Atacama Large Millimeter/submillimeter Array) era observations of molecular species that were previously difficult to observe will become commonplace. Extragalactically this will enable even high-redshift observers to move beyond simple, optically thick molecules like \co{12} to observing rarer isotopes (such as \co{13} and \CeightO), and high density gas tracers like HCN and CS \citep[e.g.][]{2014ApJ...785..149S}.  

The rare CO isotopes are especially tempting targets, as they are significantly more optically thin than \co{12}, and thus are usually thought to provide more information on the underlying molecular gas column and volume densities (e.g. Pineda et al. 2008; Wilson 2009). In order to obtain this information however, one typically needs information on the abundance of \co{13} with respect to \co{12}, the optical depth of these molecules, and the CO $X$-factor. 

The CO $X$-factor, and its variation with galaxy properties, has been discussed at length in the literature \citep[e.g][]{2011ApJ...737...12L,2013ApJ...777....5S,2013ARA&A..51..207B}, but the effect of global galaxy properties on these other two quantities has received less attention. {Some works have suggested that the \co{12} / \co{13} ratio is higher in mergers and (ultra-)luminous infrared galaxies \citep[e.g.][]{1992A&A...264...55C,1998ApJ...507L.121T,1999ApJ...522..214T,2001ApJS..135..177G}, but if this is a discrete change caused by the extreme conditions in these systems, or a continuous process has not been determined}. \co{12} / \co{13} ratios are also known to vary with galacticentric radius in extragalactic systems and in our own Milky Way \citep{1999RPPh...62..143W, 2001ApJS..135..183P}.

We use existing archival observations of \co{12}, \co{13} and \CeightO\ to show that there are systematic changes in these quantities as a function of the the star-formation rate (SFR) surface density $\Sigma_{\rm SFR}$ (or equivalently gas surface density; quantities linked through the Kennicutt-Schmidt relation: \citealt{1998ApJ...498..541K}). 
In Section \ref{results} we discuss briefly our sample selection, before presenting these results. 
We consider the cause of this effect in Section \ref{discuss}, and discuss its impact and prospects for the future in Section \ref{conclude}.

{Throughout this paper we adopt the abundances of  [$^{12}$C/$^{13}$C] =70 and [$^{16}$O/$^{18}$O] =200. These values will almost certainly change in individual objects, but the adopted values lie in the middle of the range found for extragalactic objects ([$^{12}$C/$^{13}$C]$\approx$40-130; e.g. \citealt{1993A&A...274..730H,2000A&A...358..433M,2010A&A...522A..62M} and  [$^{16}$O/$^{18}$O]$\approx$150-250; e.g. \citealt{1991A&A...249...31S,1993A&A...274..730H,1994ARA&A..32..191W}), and thus these abundances provide a reasonable standard reference. {Unless otherwise indicated all molecular line ratios discussed in this paper are intensity ratios.} }

\section{Results}
\label{results}

\subsection{Sample}

\subsubsection{Molecular data}
We collected data on the gas and star-formation properties of a sample of nearby spiral, starburst and early-type galaxies which have published \co{13} detections (see Table \ref{proptable}). We based this search on the compilation of \co{12}/\co{13} {intensity} ratios in \cite{2010MNRAS.407.2261K}, adding additional galaxies from various other sources to cover a range in star-formation and gas properties. This includes the gas rich early-type galaxies of \cite{2010MNRAS.407.2261K} and \cite{2012MNRAS.421.1298C}, but removing the Virgo cluster early-type objects which have a different behaviour, discussed in detail in Alatalo et al., in prep. 

We also checked the literature for \CeightO\ detections of the sample objects, in particular utilising the catalogue of \cite{2011RAA....11..787T}.  In this paper we show mainly correlations with the ground state {rotational} transitions of the \co{12}, \co{13} and \CeightO\ molecules. Data for the \co{13}(2-1) transitions do exist for some objects, and show the same trends. We also include this data in Table \ref{proptable}. 

\subsubsection{Star formation properties}
 The star formation rate surface densities (and gas surface densities) we use in this work are mainly taken from \cite{1998ApJ...498..541K} for the spiral and starburst galaxies ({derived using H$\alpha$ emission in spirals and far infrared emission in galaxies with circum-nuclear starbursts; \citealt{1998ApJ...498..541K}}) and \cite{2014arXiv1403.4850D} for early-type galaxies (here we use their star-formation rates derived from 22$\mu$m and far-ultraviolet emission, corrected for the effect of old stellar populations). {These objects cover a wide range of infrared luminosities (from 10$^7$-10$^{11}$ L$_{\odot}$; \citealt{1976ApJ...203L..53T, 2000A&amp;A...354..836L}), however they may not be good proxies for very luminous starburst galaxies, especially those studied at high redshift \citep[e.g.][]{2013MNRAS.436.2793D}.}
 Table \ref{proptable} lists the salient properties of the galaxies included in our sample, and the references from which the values are drawn. 
 
 {Although we use different tracers of star formation in this work, the majority of these objects are well studied nearby spiral/starbursting galaxies, in which various tracers have been observed (and indeed many of these galaxies are used as the basis for SFR calibrations).  Overall, using different tracers of star formation in these galaxies results in a scatter of $\approx$0.35 dex in the star-formation rates, but with no systematic offset (see e.g. \citealt{2002MNRAS.332..283R,2007ApJ...666..870C} for a full discussion). We thus expect that using different tracers will not unduly affect the observed relations. As a test we performed our analysis only on the subsample of galaxies where $\Sigma_{\rm SFR}$ has been estimated from far infrared emission, and found that the results stated in this paper do not change. }

\subsubsection{Aperture effects}
{We stress that the $\Sigma_{\rm SFR}$ values we use here are global, averaged over the entire galaxy disk. However, the \co{12}, \co{13} and \CeightO\ detections are from single dish telescope observations toward galaxy centres. In general these observations had a large beam size ($\approx$22\arcsec\ at 3mm for the majority of these observations that were taken with the IRAM-30m telescope) so that the measured line emission arises from the majority of the galaxies star-forming disk (the inner $\ltsimeq$10 kpc, depending on distance). The millimetre line ratios and SFR should thus be comparable (as both the star formation and gas in galaxies are centrally concentrated). However, as formally these quantities are not measured in the same apertures this could increase the scatter on correlations determined from these data; for instance if radial gradients are present. We check for such effects in Section \ref{radgrad}, and do not see strong evidence for any bias. 
Ideally, however, one would extend this work using resolved maps of star-formation rate densities and gas properties to formally match these quantities, but that is outside the scope of this first paper. }

\begin{table*}
\caption{Literature values used in this work.}
\begin{tabular*}{0.9\textwidth}{@{\extracolsep{\fill}}l r r r r r r r r r}
\hline
Name & $\Sigma_{\rm SFR}$ & Tracer & Ref. & \co{12}(1-0)/\co{13}(1-0) & Ref. & \co{12}(2-1)/\co{13}(2-1) & Ref. &  \co{12}/\CeightO\ & Ref \\
 (1) & (2) & (3) & (4) & (5) & (6) &(7) & (8) & (9) & (10) \\
 \hline
Circinus & 0.80 & 24$\mu$m & 11 & 10.1 $\pm$ 1.4 & 13 &  -- & --  & 54.4 $\pm$ 5.4 & 15\\
IC0676 & -0.46 & 22$\mu$m+FUV & 2 & 7.5 $\pm$ 1.2 & 13 & 8.3 $\pm$ 1.3 & 13 &  --  &  -- \\
IC1024 & -0.93 & 22$\mu$m+FUV & 2 & 15.1 $\pm$ 2.8 & 13 & 8.1 $\pm$ 1.2 & 13 &  --  &  -- \\
IC342 & -0.41 & FIR & 1 & 8.4 $\pm$ 0.5 & 13 &  -- & --  & 35.1 $\pm$ 3.0 & 15\\
Maffei2 & -2.36 & FIR & 12 & 8.4 $\pm$ 1.0 & 16 &  -- & --  & 31.6 $\pm$ 5.4 & 18\\
NGC0891 & -2.39 & 24$\mu$m & 6 & 5.0 $\pm$ 0.5 & 13 &  -- & --  &  --  &  -- \\
NGC1068 & 1.80 & 8$\mu$m & 20 & 13.5 $\pm$ 3.7 & 13 & 16.7 $\pm$ 2.8 & 13 & 47.8 $\pm$ 12.7 & 16\\
NGC1222 & 0.55 & 22$\mu$m+FUV & 2 & 21.1 $\pm$ 4.1 & 13 & 20.0 $\pm$ 4.1 & 13 &  --  &  -- \\
NGC1266 & 1.73 & 22$\mu$m+FUV & 2 & 30.9 $\pm$ 5.3 & 13 & 24.8 $\pm$ 3.7 & 13 &  --  &  -- \\
NGC1808 & 0.08 & FIR & 1 & 16.7 $\pm$ 2.8 & 13 & 14.3 $\pm$ 2.0 & 13 & 48.9 $\pm$ 3.3 & 15\\
NGC2146 & 0.84 & FIR & 1 & 12.5 $\pm$ 1.6 & 13 & 10.0 $\pm$ 1.0 & 13 & 32.7 $\pm$ 9.1 & 15\\
NGC253 & 1.24 & FIR & 1 & 10.0 $\pm$ 1.0 & 13 & 12.0 $\pm$ 1.2 & 19 & 61.0 $\pm$ 6.2 & 15\\
NGC2764 & -0.82 & 22$\mu$m+FUV & 2 & 10.8 $\pm$ 1.7 & 13 & 11.6 $\pm$ 1.8 & 13 &  --  &  -- \\
NGC3032 & -1.26 & 22$\mu$m+FUV & 2 & 9.5 $\pm$ 1.5 & 14 & 5.3 $\pm$ 0.8 & 14 &  --  &  -- \\
NGC3034 & 1.48 & FIR & 1 & 14.3 $\pm$ 1.4 & 13 & 10.0 $\pm$ 1.0 & 13 & 54.5 $\pm$ 13.0 & 15\\
NGC3079 & 1.63 & FIR & 1 & 16.7 $\pm$ 2.8 & 13 & 14.3 $\pm$ 2.0 & 13 &  --  &  -- \\
NGC3184 & -2.76 & FIR & 10 & 4.9 $\pm$ 1.2 & 13 &  -- & --  &  --  &  -- \\
NGC3256 & 0.68 & FIR & 1 & 33.3 $\pm$ 11.1 & 13 & 10.0 $\pm$ 4.0 & 13 & 91.6 $\pm$ 28.1 & 15\\
NGC3556 & -2.38 & Radio. Cont. & 7 & 8.7 $\pm$ 1.4 & 13 &  -- & --  &  --  &  -- \\
NGC3593 & -0.17 & FIR & 8 & 10.6 $\pm$ 2.2 & 13 &  -- & --  &  --  &  -- \\
NGC3607 & -2.22 & 22$\mu$m+FUV & 2 & 5.9 $\pm$ 1.0 & 13 & 6.4 $\pm$ 1.3 & 13 &  --  &  -- \\
NGC3627 & -0.77 & FIR & 1 & 16.7 $\pm$ 1.7 & 13 &  -- & --  &  --  &  -- \\
NGC3628 & -2.73 & FIR & 10 & 8.3 $\pm$ 1.4 & 13 & 10.0 $\pm$ 2.0 & 13 &  --  &  -- \\
NGC3665 & -2.34 & 22$\mu$m+FUV & 2 & 3.0 $\pm$ 0.5 & 13 & 2.4 $\pm$ 0.4 & 13 &  --  &  -- \\
NGC3982 & -1.19 & H$\alpha$ & 3 & 14.3 $\pm$ 2.0 & 13 &  -- & --  &  --  &  -- \\
NGC4150 & -1.41 & 22$\mu$m+FUV & 2 & 13.7 $\pm$ 3.2 & 14 & 12.2 $\pm$ 2.4 & 14 &  --  &  -- \\
NGC4258 & -2.36 & H$\alpha$ & 1 & 10.0 $\pm$ 1.0 & 13 &  -- & --  &  --  &  -- \\
NGC4826 & -2.47 & H$\alpha$ & 1 & 8.3 $\pm$ 1.4 & 13 & 7.7 $\pm$ 1.2 & 13 & 19.0 $\pm$ 2.7 & 15\\
NGC4945 & 1.22 & FIR & 9 & 16.7 $\pm$ 2.8 & 13 & 9.1 $\pm$ 0.8 & 13 &  --  &  -- \\
NGC5033 & -2.64 & H$\alpha$ & 1 & 8.3 $\pm$ 0.7 & 13 &  -- & --  &  --  &  -- \\
NGC5055 & -2.32 & H$\alpha$ & 1 & 5.7 $\pm$ 0.6 & 13 &  -- & --  &  --  &  -- \\
NGC5194 & -1.78 & FIR & 1 & 6.7 $\pm$ 0.7 & 13 & 10.0 $\pm$ 1.0 & 13 & 29.1 $\pm$ 10.3 & 15\\
NGC5236 & -1.41 & FIR & 1 & 10.0 $\pm$ 1.0 & 13 & 10.0 $\pm$ 3.0 & 13 &  --  &  -- \\
NGC5866 & -0.64 & 22$\mu$m+FUV & 2 & 6.2 $\pm$ 0.9 & 13 & 4.3 $\pm$ 0.7 & 13 &  --  &  -- \\
NGC6014 & -2.16 & 22$\mu$m+FUV & 2 & 9.0 $\pm$ 2.0 & 13 & 6.7 $\pm$ 1.1 & 13 &  --  &  -- \\
NGC6946 & -1.88 & FIR & 1 & 13.3 $\pm$ 1.3 & 13 &  -- & --  & 21.0 $\pm$ 3.0 & 17\\
NGC7172 & -2.40 & H$\alpha$ & 4 & 6.7 $\pm$ 0.7 & 13 &  -- & --  &  --  &  -- \\
NGC7331 & -2.33 & H$\alpha$ & 1 & 7.1 $\pm$ 0.7 & 13 & 5.9 $\pm$ 1.0 & 13 &  --  &  -- \\
NGC7469 & 0.66 & FIR & 5 & 16.7 $\pm$ 1.7 & 13 & 20.0 $\pm$ 2.0 & 13 &  --  &  -- \\
NGC7552 & 0.16 & FIR & 1 & 14.3 $\pm$ 2.0 & 13 & 9.1 $\pm$ 1.7 & 13 &  --  &  -- \\
PGC058114 & -0.16 & 22$\mu$m+FUV & 2 & 18.9 $\pm$ 4.1 & 14 & 14.4 $\pm$ 2.8 & 14 &  --  &  -- \\
    \hline
\end{tabular*}
\parbox[t]{0.9 \textwidth}{ \textit{Notes:} Column 1 contains the names of the galaxies considered in this work. Column 2 contains the base-10 logarithm of their globally integrated surface density of star formation in units of M$_{\odot}$ yr$^{-1}$ kpc$^{-2}$. Column 3 and 4 contain the SFR tracer, and the references, respectively. The \co{12}(1-0)/\co{13}(1-0) flux density ratio is given in Column 5, along with its associated error, while Column 6 lists the reference these were taken from. Columns 7 and 8 similarly show the \co{12}(2-1)/\co{13}(2-1) values and references, while Columns 9 and 10 show the \co{12}/\CeightO\ ratios with errors, and references. References; 1: \cite{1998ApJ...498..541K}, 2: \cite{2014arXiv1403.4850D}, 3: \cite{2008ChJAA...8..555Z}, 4: \cite{2005AJ....129.2597H}, 5: \cite{2000ApJ...537..631P}, 6: \cite{2011AJ....141...48Y}, 7: \cite{1997NewA....2..251K}, 8: \cite{2000A&A...363..869G}, 9: \cite{1993A&A...270...29D}, 10: \cite{2000A&amp;A...354..836L}, 11: \cite{2012MNRAS.425.1934F}, 12: \cite{1994ApJ...421..122T}, 13: \cite{2010MNRAS.407.2261K}, 14: \cite{2012MNRAS.421.1298C}, 15: \cite{2011RAA....11..787T}, 16: \cite{2013A&A...549A..39A}, 17: \cite{2004AJ....127.2069M}, 18: \cite{2008ApJ...675..281M}, 19: \cite{2014A&amp;A...564A.126R}. 20:\cite{2012ApJ...746..129T}. }
\label{proptable}
\end{table*}

\subsection{\co{12} / \co{13} as a function of $\Sigma_{\rm SFR}$}

 \begin{figure}
\begin{center}
\includegraphics[width=0.45\textwidth,angle=0,clip,trim=0.0cm 1.5cm 0cm 0.0cm]{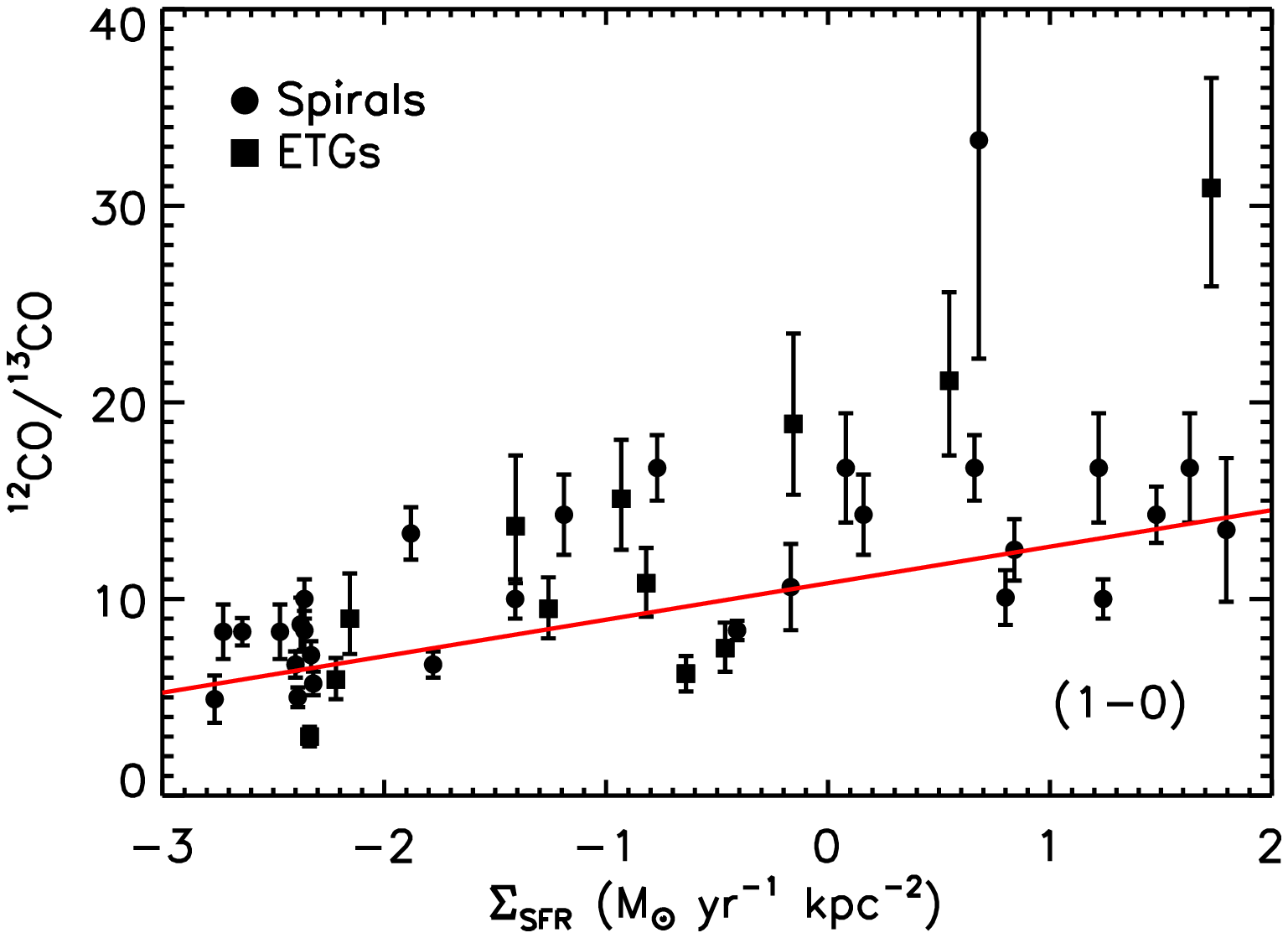}
\includegraphics[width=0.45\textwidth,angle=0,clip,trim=0.0cm 0.0cm 0cm 0.0cm]{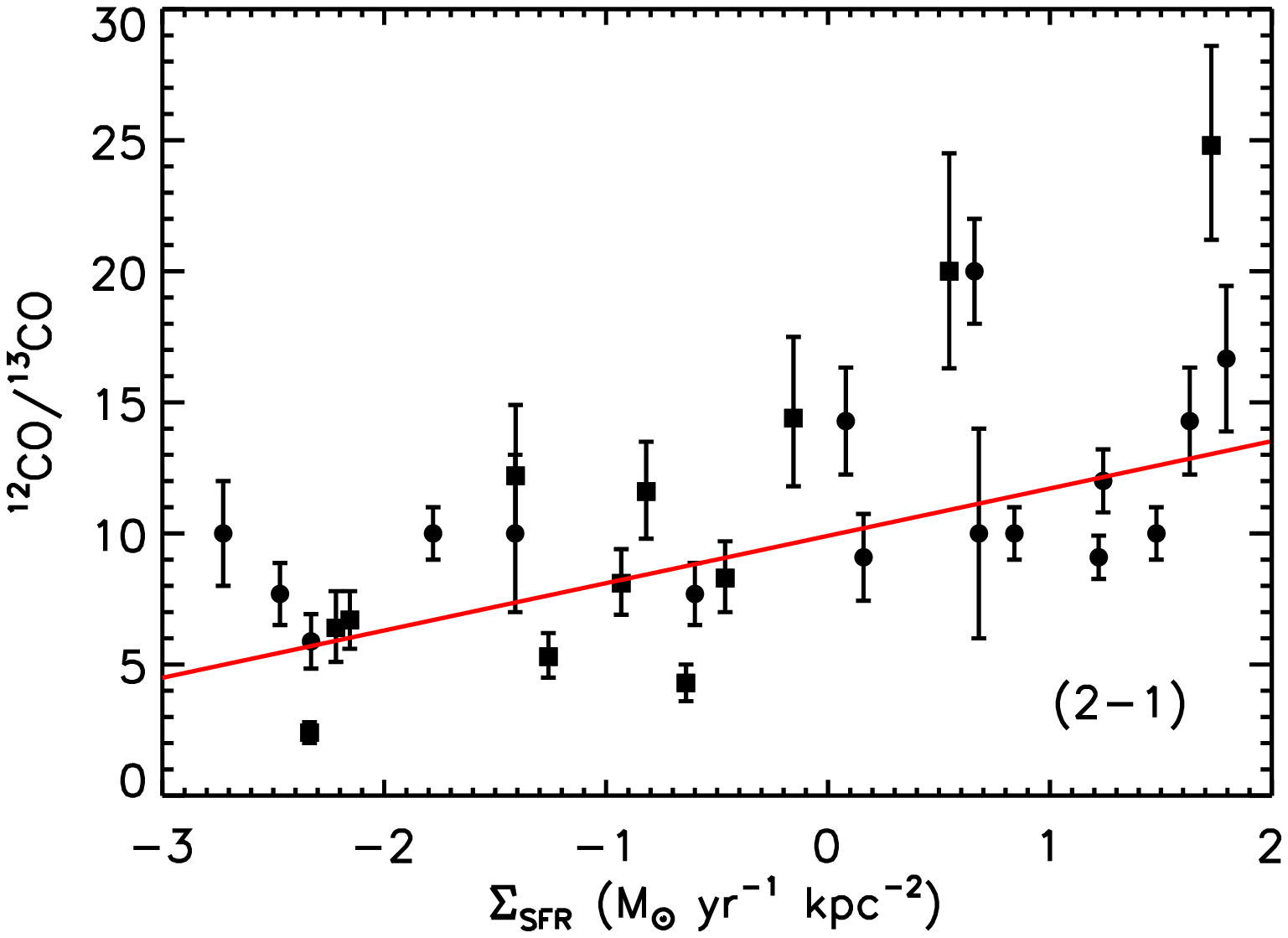}
 \end{center}
\caption{The \co{12} / \co{13}  {intensity} ratio for our sample galaxies, plotted against their average star-formation rate surface density. The top panel shows the 1-0 transitions, and the bottom panel the 2-1 transitions. Black circles show spiral and starburst galaxies, while black squares denote early-type galaxies. The red line shows a best fit to the data, as outlined in the text.  }
\label{13coratio}
 \end{figure}

Using the literature data from Table \ref{proptable}, following other authors we do not find a dependance of the \co{12} / \co{13}  {intensity} ratio with total star formation rate. However, as shown in Figure \ref{13coratio} we report for the first time a strong positive correlation (Spearmans rank correlation coefficient of 0.755 for the ground transition) between the \co{12} / \co{13}  {intensity} ratio and $\Sigma_{\rm SFR}$. We show on Figure \ref{13coratio} the best linear fit to this correlation (plotted as a red line), finding that 

\begin{equation}
\frac{\rm ^{12}CO(\mbox{1--0})}{\rm ^{13}CO(\mbox{1--0})} = (1.84\pm0.14) \log_{10}({\Sigma_{\rm SFR}}) + (10.80 \pm 0.28),
\label{r10eq}
\end{equation}
\noindent and
\begin{equation}
\frac{\rm ^{12}CO(\mbox{2--1})}{\rm ^{13}CO(\mbox{2--1})} = (1.80\pm0.31) \log_{10}({\Sigma_{\rm SFR}}) + (9.91 \pm 0.51).
\label{r21eq}
\end{equation}

Thus within errors both transitions follow the same trend. 
{We warn however that these fits were conducted using standard least-squares techniques, with points weighted by the formal error bars shown in the plots. The error bars in the  \co{12} / \co{13} ratios are in reality non-gaussian (when estimated from low signal-to-noise data; see \citealt{2012MNRAS.421.1298C} for a full discussion). Correcting the literature data used in this paper for this effect is beyond the scope of this work, and thus the exact values of the slope and intercept in these fits should be treated with caution. The effect of non-gaussian errors should be considered in future studies of this phenomenon.}

Throughout this work the reported correlations are equally strong when plotted as a function of $\Sigma_{\rm SFR}$ or gas surface density (as expected due to these parameters being linked in the Schmidt-Kennicutt Relation; e.g. \citealt{1998ApJ...498..541K}). Which of these parameters drives the effect (or indeed if either parameter does) will be  discussed in Section \ref{discuss}.

In Figure \ref{13coratio} the spiral and starburst galaxies are shown with black circles, and the early-type objects with black squares. No significant difference between the two populations is detectable. Nuclear activity also appears to matter little, with Seyfert, starburst and star-forming nuclei (as defined in \citealt{2010MNRAS.407.2261K}) all following the same mean trend. Two galaxies do appear to have significantly higher \co{12} / \co{13} ratios then the mean population. Interestingly these objects (NGC1266 and NGC3256) both have nuclear molecular outflows (\citealt{2011ApJ...735...88A,2014arXiv1403.7117S}). {M82 (NGC3034) which has a famous starburst driven outflow is also present in this work, but follows the same mean trend as the normal galaxies. The outflowing gas is at large radii in this object, and this may explain why we do not detect any differences in the \co{12} / \co{13} ratio within the nuclear region.}

 \begin{figure}
\begin{center}
\includegraphics[width=0.45\textwidth,angle=0,clip,trim=0.0cm 0.0cm 0cm 0.0cm]{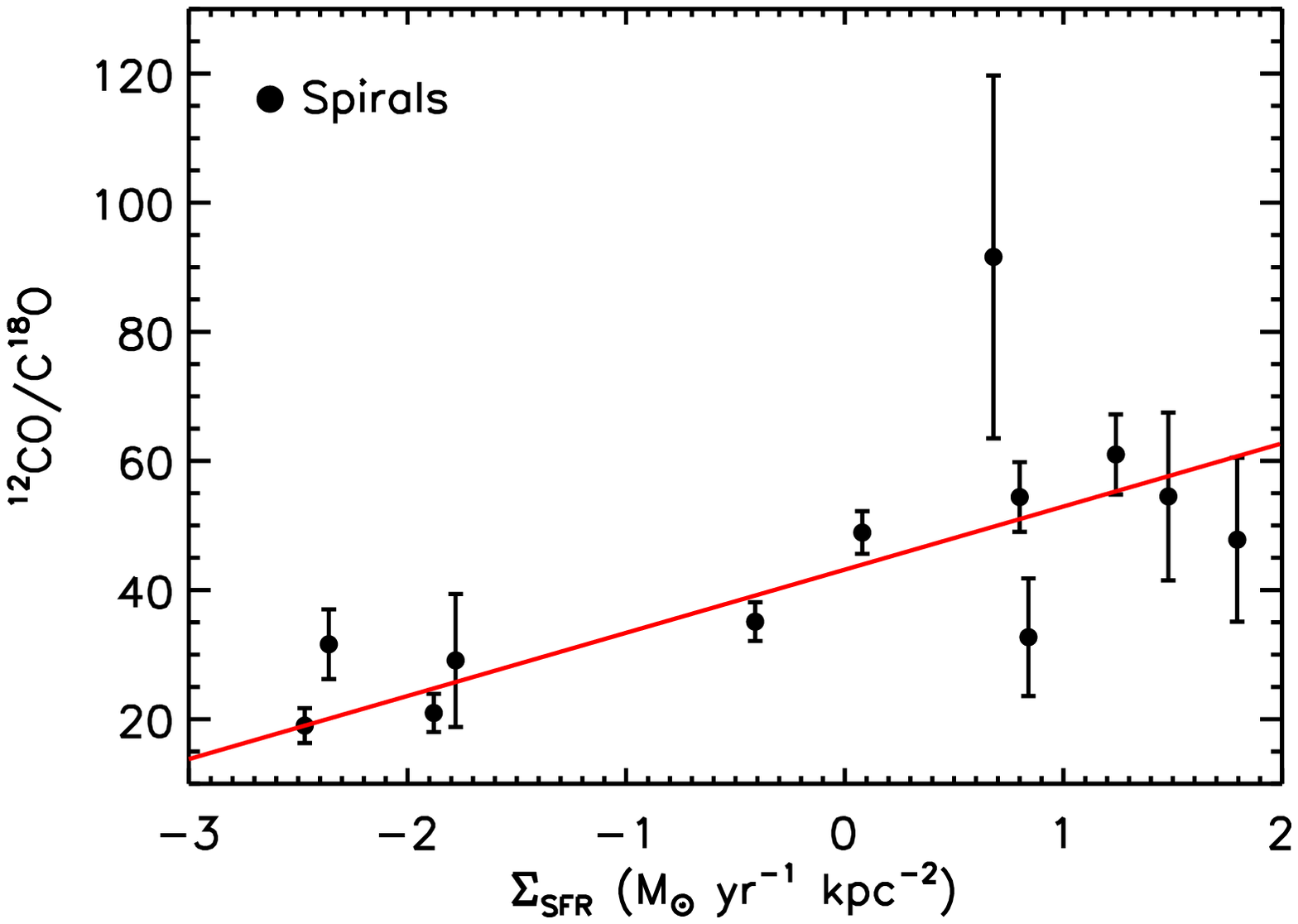}
 \end{center}
\caption{As Figure \protect \ref{13coratio}, but showing the \co{12} / \CeightO\ ratio as a function of $\Sigma_{\rm SFR}$. }
\label{c18ofig}
 \end{figure}

\subsection{Correlations with \CeightO}
\label{c8ocorrs}
In Table \ref{proptable} we also tabulate \CeightO\ line {intensity} ratios for objects where this data exists. \CeightO\ is a rare molecule, ensuring it is almost always optically thin. It also is produced via different {nucleosynthetic} pathways than \co{13}, giving it substantial power to break degeneracies present in studies of only the \co{12}/\co{13} ratio. In Figure \ref{c18ofig} we show that the \co{12}/\CeightO\ ratio also shows a positive correlation with $\Sigma_{\rm SFR}$ (once again the outlier is molecular outflow galaxy NGC3256). We again show the best linear fit as a red line, finding that 

\begin{equation}
\frac{\rm ^{12}CO(\mbox{1--0})}{\rm C^{18}O(\mbox{1--0})} = (9.8\pm1.0) \log_{10}({\Sigma_{\rm SFR}}) + (43.2 \pm 1.7).
\label{c18oeq}
\end{equation}

\begin{figure}
\begin{center}
\includegraphics[width=0.45\textwidth,angle=0,clip,trim=0.0cm 0.0cm 0cm 0.0cm]{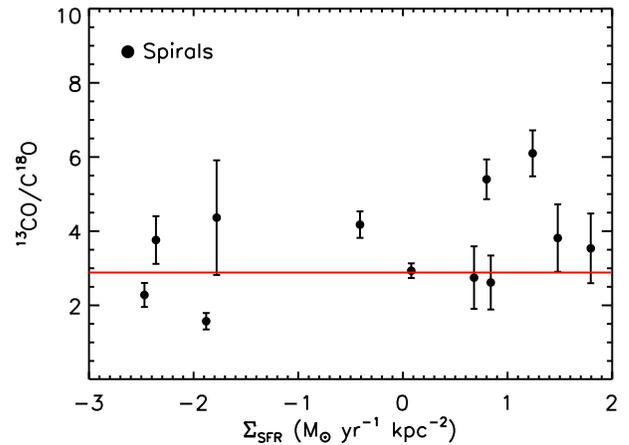}
 \end{center}
\caption{As Figure \protect \ref{13coratio}, but showing the \co{13} / \CeightO\ ratio as a function of $\Sigma_{\rm SFR}$. The red line is a fit to the data with a fixed (flat) slope, which is consistent with the relation expected for a constant \co{13} / \CeightO\ fractional abundance of 2.89 (with our adopted abundances), if both transitions are optically thin.}
\label{c18oto13co}
 \end{figure}

In Figure \ref{c18oto13co} we show the ratio of \co{13} to \CeightO, plotted against $\Sigma_{\rm SFR}$. The line ratio is consistent with being flat over 5 orders of magnitude in $\Sigma_{\rm SFR}$, with a fitted intercept of 2.89 (despite the small number of detected objects and large observational errors). The observed line ratios are all consistent with the simple ratio between our adopted abundances of [\co{13}/\CeightO] = 200/70 = 2.86.  This suggests both that \co{13} is not becoming significantly optically thick in these objects, and that the relative fractional abundance of these molecules does not change as a function of $\Sigma_{\rm SFR}$.

\section{Discussion}
\label{discuss}

In the previous Section we have shown that the \co{12}/\co{13}  {intensity} ratio varies systematically as a function of $\Sigma_{\rm SFR}$ (Figure \ref{13coratio}). The same trend is observed in the \co{12}/\CeightO\ ratio (Figure \ref{c18ofig}). Galaxies of different types follow the same trend, and the trend is not found to alter in objects with significant nuclear activity. Galaxies with {nuclear} molecular outflows, however, do tend to lie above the relation. We below discuss the potential physical drivers of the observed correlations.

{Previous studies have shown that the \co{12} / \co{13} intensity ratio is higher in (ultra-)luminous infrared galaxies \citep[e.g.][]{1992A&A...264...55C,2001ApJS..135..177G}, and that this line ratio appears to correlate with total star-formation rate \citep[or a proxy such a far-infrared luminosity; e.g.][]{1998ApJ...507L.121T,1999ApJ...522..214T}. We show here for the first time that this effect can be observed in galaxies of all types, down to much lower total star-formation rates, and appears to be linked to $\Sigma_{\rm SFR}$ rather than the total star-formation rate itself. ULIRGs tend to be very compact star-forming systems at low redshift, and so for them a change in total star-formation rate likely also indicates a change in $\Sigma_{\rm SFR}$. The same is not true in spiral and early-type galaxies, and thus a strong correlation does not exist between total SFR and the \co{12} / \co{13} intensity ratio in our study.}

\subsection{Cause of the observed variations}
\label{causevar}
In order to produce the changes in the \co{12}/\co{13} ratio we observe:

\begin{enumerate}
\item the lines could be excited differently. However, as we are observing low-$J$ CO lines, and we see the same relation for both the $J$=1--0 and $J$=2--1 transitions, we consider this unlikely, and do not discuss this possibility further here.
\item there is a change in the fractional abundance of  \co{13} and \CeightO\ with respect to \co{12}. The fractional abundance of \co{13} and \CeightO\ can be affected by selective photodissociation, isotope dependant fractionation and selective nucleosynthesis. We discuss each of these processes in detail below.
\item the optical depth of the gas is changing. In order to reproduce the observed effect either \co{12} or \co{13} could be becoming more optically thin in high $\Sigma_{\rm SFR}$ galaxies. One can distinguish between the possibilities using \CeightO, which has a sufficiently low abundance to be almost always optically thin. As shown in Figure \ref{c18oto13co} the ratio of \co{13}/\CeightO\ does not change as a function of $\Sigma_{\rm SFR}$, as would be expected if the \co{13} optical depth was changing. We therefore here suggest that (if optical depth changes are causing this correlation) this is caused by a change in the optical depth of \co{12}. This possibility will be discussed further in Section \ref{optdepthchanges}.
\end{enumerate}

\subsubsection{Selective photodissociation of $^{13}$CO and C$^{18}$O}

In a strong starburst OB stars emit large amounts of light in the ultraviolet, generating a strong interstellar  radiation field. This emission penetrates some distance into molecular clouds, but is unable to  photodissociate substantial amounts of \co{12} due to shielding effects \citep{1988ApJ...334..771V}. \co{13} however is not as protected, and will be preferentially destroyed as the UV radiation field increases in strength. This could naturally lead to the observed correlation, as galaxies with high SFR surface densities will form more OB stars per unit area and thus have a stronger interstellar radiation field. However, \CeightO\ is dissociated three times faster than \co{13} ({due to a combination of lower self shielding and differences in its molecular structure; \citealt{1988ApJ...334..771V}}), inconsistent with the apparent constant \co{13}/\CeightO\  {intensity} ratios in Figure \ref{c18oto13co}.  This suggests selective photodissociation is not the dominant effect driving the correlation in Figure \ref{13coratio}.
 
\subsubsection{Isotope dependant fractionation} 

In cold regions ($\approx$10 K) the rapid exothermic exchange of the $^{13}$C$^+$ ion with $^{12}$C in \co{12} preferentially forms \co{13} \citep[e.g.][]{1976ApJ...205L.165W,1998ApJ...494L.107K}.  In regions with higher temperatures, however, rapid ion exchange would negate the effects of this process.  The decrease in \co{12} abundance would not be noticeable, due to the large optical depth of this transition, but the  \co{12}/ \co{13}  {intensity} ratio would be observed to change, as the additional \co{13} would emit more strongly.

If the mean temperature of the clouds in our galaxies increases as a function of $\Sigma_{\rm SFR}$ (as might be expected, due to the increased contribution of heating from newly formed stars) this could create a correlation that resembles that seen in Figure \ref{13coratio}. Whether the cloud temperatures in the disks of our galaxies are really low enough to support large-scale fractionation is difficult to determine. However, this fractionation process would change the \co{13} abundance with respect to \CeightO\ (which is unaffected by fractionation), inconsistent with the apparent constant \co{13}/\CeightO\  {intensity} ratios in Figure \ref{c18oto13co}.

If the $\Sigma_{\rm SFR}$ does correlate with the mean temperature in the clouds, then it is also possible that local freeze out of CO onto the ices coating dust grains could alter the fractional abundances in a systematic way. However, freeze out is not expected to effect galaxy scale abundances, as the fraction of gas as ice is negligible, even in cold Galactic systems.

 \subsubsection{Selective nucleosynthesis of $^{12}$C}
 
 Massive stars that end their lives as supernovae produce $^{12}$C, but do not produce significant amounts of $^{13}$C, which is instead produced primarily by low mass stars, and injected into the ISM from AGB star winds \citep[e.g.][]{1991A&A...249...31S}. Thus one could explain the positive correlation we see between the \co{12}/\co{13}  {intensity} ratio and $\Sigma_{\rm SFR}$ as a timescale effect, where selective nucleosynthesis of $^{12}$C in highly star-forming regions changes the $^{12}$C/$^{13}$C abundance \citep[e.g.][]{1999RPPh...62..143W}.
 However, $^{18}$O is primarily produced by massive stars \citep{1991A&A...249...31S,1995ApJS...98..617T,rmg.2008.68.4,2013ARA&A..51..457N}, and hence selective nucleosynthesis can not explain the positive change in the \co{12}/\CeightO\  {intensity} ratio with SFR density (Figure \ref{c18ofig}), or the lack of a correlation seen in the \co{13}/\CeightO\ {intensity} ratio (Figure \ref{c18oto13co}).
 
\subsection{Physical drivers for optical depth changes}
\label{optdepthchanges}

\begin{figure}
\begin{center}
\includegraphics[width=0.45\textwidth,angle=0,clip,trim=-0.1cm 0.0cm 0cm 0.0cm]{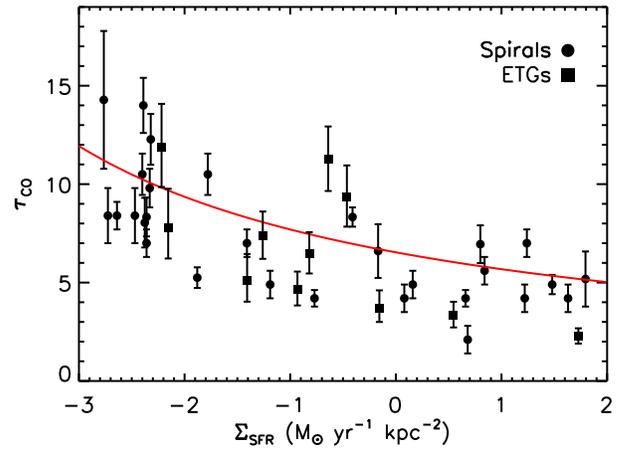}
 \end{center}
\caption{The mean optical depth of the \co{12} gas in our sample galaxies, derived from Equation \protect \ref{taueq} (assuming abundances as discussed above), plotted against $\Sigma_{\rm SFR}$. The red curve is the best fit from Figure \protect \ref{13coratio}, converted into an optical depth using Equation \protect \ref{taueq}.}
\label{13coratiotau}
 \end{figure}

As discussed above, we can plausibly eliminate the various physical processes that could result in abundance variations and cause the observed correlation. The most convincing remaining explanation is that the mean optical depth is changing in these galaxies. Assuming that the \co{13} line is optically thin, and the \co{12} line is optically thick, their ratio is proportional to the optical depth of the \co{12} gas, with the proportionality constant being the fractional abundance of $^{12}$C with respect to $^{13}$C. 
In Figure \ref{13coratiotau} we plot the optical depth of \co{12}  (here denoted $\tau_{\rm CO}$) derived from the \co{12} / \co{13}  {intensity} ratio (R$_{10}$), using the formula

\begin{equation}
\tau_{\rm CO} = \frac{[\mathrm{^{12}C/^{13}C}]}{R_{10}}.
\label{taueq}
\end{equation}

Here we assume the standard abundances of  [$^{12}$C/$^{13}$C], as discussed {in Sections \ref{intro} and \ref{c8ocorrs}}. 
The trend line from Figure \ref{13coratio} is also shown in this Figure, after conversion with Equation \ref{taueq}. If a changing optical depth is the correct interpretation, then the mean optical depth of \co{12} is up to $\approx$2.5 times lower in the most star forming objects (that have high $\Sigma_{\rm SFR}$).

As discussed in \cite{2001ApJS..135..183P}, in local thermodynamic equilibrium the \co{12} optical depth ($\tau$) varies as $\tau$ $\propto$ $N_c$/($\sigma$ $T_{k}^{2}$), where $N_c$ is the gas surface density of the cloud, $\sigma$ is the gas velocity dispersion and $T_k$ is the gas kinetic temperature (this in the limit where $T_k$ is large compared to the background temperature). 

To first order from this equation one would predict increasing optical depth with increasing gas surface density from this equation, the opposite to the effect observed. Thus we suggest that it is the gas velocity dispersion and/or kinetic temperature driving the observed effect.
Theoretically, a similar correlation between the mean optical depth of low-$J$ CO lines (within resolved GMC's) and the star-formation rate surface density has been shown by \citet[][Fig 5]{2014arXiv1401.2998N}. Although the normalisation is somewhat different, that a similar trend exists in chemical models backs up the analysis presented here.

As discussed above, the proposed change in gas velocity dispersion/temperature correlates with both $\Sigma_{\rm SFR}$ and the gas surface density. Intuitively, one might expect higher density gas to have a lower temperature (as cooling is enhanced at higher densities), and have a lower velocity dispersion (as turbulence is damped more quickly). This is the opposite to the trend observed. However, in regions with higher star-formation surface density a larger number of massive OB stars are packed into the same area. These stars have strong ionising radiation fields, drive powerful stellar winds, and explode as supernovae, all of which could increase the gas velocity dispersion and temperature.  Physically, we thus suspect a change in the gas velocity dispersion/temperature could be driven by the action of massive stars (motivating our decision to plot the correlations in this work as a function of $\Sigma_{\rm SFR}$). Determining if it is really the gas velocity dispersion, temperature, or a combination of the two which changes in our objects is, however, beyond the scope of this work.
 
 \subsection{Radial gradients}
 \label{radgrad}
 {\cite{2001ApJS..135..183P} and \cite{2011RAA....11..787T} report changes in the \co{12}/\co{13}  {intensity} ratio as a function of radius within galaxies. 
\cite{2001ApJS..135..183P} and \cite{2011RAA....11..787T} also concluded that radial temperature and/or velocity dispersion changes cause these variations in their sample of spiral galaxies, but did not connect this to the star-formation or gas surface densities. As $\Sigma_{\rm SFR}$ generally increases towards galaxy centres, the trend reported here could explain these results. }
 
{In this work we have plotted integrated star formation properties against millimetre line ratios derived from single dish telescope pointings towards galaxy centres. As these quantities are not derived in matched apertures, we could be systematically biased in nearby objects, especially if strong radial gradients (such as those discussed above) are present. We checked for any strong effect by plotting the residuals of the correlations shown above (e.g. the residuals around Equation \ref{r10eq}) against distance. As the majority of these observations were taken with the same telescope (and thus with the {same beam size}), the distance linearly correlates with the linear size of the region probed. No correlation was found, suggesting that we are not strongly biased by aperture effects.  A further, in depth analysis of the resolved effect of $\Sigma_{\rm SFR}$ on molecular line ratios within galaxies will be conducted in a future work. }

\section{Conclusions}
\label{conclude}
In this paper we have shown that the \co{12}/\co{13}  {intensity} ratio in galaxies varies systematically as a function of $\Sigma_{\rm SFR}$ and gas surface density (Figure \ref{13coratio}). The same effect is observed in different transitions, and in the \co{12}/\CeightO\ ratio, while the \co{13}/\CeightO\ ratio appears to remain constant as a function of $\Sigma_{\rm SFR}$.

We discussed the cause of these variations, considering both changes in the physical state of the gas, and chemical changes that lead to abundance variations. We used the observed correlations with \CeightO\ to suggest that abundance variations are unlikely to be causing the systematic trend observed with $\Sigma_{\rm SFR}$, and thus that the mean gas temperature and/or velocity dispersion may be systematically higher in higher star-formation rate surface density regions. We propose this effect could be caused by the combined action of massive stars heating and/or inducing turbulence in the gas phase.

We present the best fitting relations between $\Sigma_{\rm SFR}$ and the \co{12}/\co{13} and \co{12}/\CeightO\  {intensity} ratios, in the hope they will prove useful to the community. For instance, \cite{2014arXiv1402.1456C} have shown that X$_{\rm CO}$ also appears to vary with $\Sigma_{\rm SFR}$ (their Figure 43), leading to the intriguing possibility that these results may be linked. In addition, \cite{2014arXiv1401.2998N} showed it is possible to estimate the expected CO spectral line energy distributions using $\Sigma_{\rm SFR}$. By combining with the relations presented here it is also possible to determine the expected brightness of \mbox{$\mathrm{low-}J$} $^{13}$CO and \CeightO\ lines for galaxies across the universe. 

As telescopes such as ALMA come online we will gain access to CO line ratios in high-$z$ sources that are not easily spatially resolved. If this relation were found to hold in such systems, it could be inverted to estimate $\Sigma_{\rm SFR}$ (and with knowledge of the total star-formation rate, give some idea of the size of the star-forming disk). In addition, we highlight that galaxies with molecular outflows tend to lie above the relations presented here (likely due to enhanced emission from their optically thin outflows). This relation could thus be useful to identify potential outflow candidates for further followup.

Although we showed here that the gas velocity dispersion and/or temperature may change systematically in galaxies with high $\Sigma_{\rm SFR}$, determining which of these variables is really responsible will require additional work. For instance, high resolution observations which resolve individual molecular clouds in various CO isotopologues can be used to both estimate the velocity dispersion and temperature of the gas (through non-thermodynamic equilibrium chemical modelling). Comparing to Galactic clouds could also shed light on this effect, and help us predict the expected emission from galaxies across the known universe. 

\vspace{0.5cm}
\noindent \textbf{Acknowledgments}

TAD acknowledges the support provided by an STFC Ernest Rutherford Fellowship, and thanks Izaskun Jimenez-Serra, Adam Ginsburg, Andreas Schruba and Diederik Kruijssen for useful discussions. 
The research leading to these results has received funding from the European
Community's Seventh Framework Programme (/FP7/2007-2013/) under grant agreement
No 229517.

\bsp
\bibliographystyle{mn2e}
\bibliography{13COpaper.bib}
\bibdata{13COpaper.bib}
\bibstyle{mn2e}

\label{lastpage}
\end{document}